# Autoplot: A browser for scientific data on the web

J. Faden, R. S. Weigel, J. Merka, R. H. W. Friedel

# Abstract

Autoplot is software developed for the Virtual Observatories in Heliophysics to provide intelligent and automated plotting capabilities for many typical data products that are stored in a variety of file formats or databases. Autoplot has proven to be a flexible tool for exploring, accessing, and viewing data resources as typically found on the web, usually in the form of a directory containing data files with multiple parameters contained in each file. Data from a data source is abstracted into a common internal data model called QDataSet. Autoplot is built from individually useful components, and can be extended and reused to create specialized data handling and analysis applications and is being used in a variety of science visualization and analysis applications. Although originally developed for viewing heliophysics-related time series and spectrograms, its flexible and generic data representation model makes it potentially useful for the Earth sciences.

# 1. Introduction

Autoplot was first developed for the ViRBO (Virtual Radiation Belt Observatory; http://virbo.org/) to provide its plotting capabilities, and has since been generalized into a flexible software framework that is used in a variety of applications by several virtual observatories and satellite instrument groups. Autoplot can read data from a variety of sources, such as CDF (http://cdf.gsfc.nasa.gov/), ASCII, OPeNDAP (Gallagher et al, 2008; http://opendap.org/), CEF (Cluster Exchange Format), Excel spreadsheets, and NetCDF (Fulker, 1988; http://www.unidata.ucar.edu/software/netcdf) into a common internal data representation. A fundamental concept is that datasets are identified by compact URIs (Uniform Resource Identifiers).

Autoplot is designed to enable a variety of data analysis and visualization applications. There are many different application-specific software applications for plotting time series and spectrogram-type scientific data in heliophysics. Examples of Java-based software include ViSBARD (Visual System for Analysis and Retrieval of Data; http://spdf.gsfc.nasa.gov/research/visualization/visbard/), and DataShop (Vandegriff et al., 2004;2009). An example from the Earth Sciences is IDV (Interactive Data Viewer; http://www.unidata.ucar.edu/software/idv/).

All of these applications support basic views of time series and spectrogram data, among many other features. Based on our experience derived from development of plotting packages with similar functionality, (PaPCo [http://papco.org/], for Panel Plot Composer, a predecessor of Autoplot, and Das2 [http://das2.org/]), implementing a basic view of time series and a spectrogram at the level of most existing software packages takes little time in comparison to the alternative of learning how to use an existing plotting package.  However, there are many complexities that must be handled in order for the application to be able to render many of the time series and spectrograms found on the web.  These complexities include missing data, various time representations, multiple fill values, and various metadata representations.  Much of the development effort for Autoplot has been in developing data handlers that gracefully handle these complexities, which are typically not well handled by existing software packages (without extensive additional development). The initial objective for Autoplot was that given the URI to a scientific data file, which may or may not conform to a given specification, it should automatically render a view of the data in the file with reasonable axis labels and guesses for the title based on any available metadata.

To support this, an internal representation of data, QDataSet (for "Quick Dataset"), described in Section 4, was developed that has much overlap with that of NetCDF (http://my.unidata.ucar.edu/content/software/netcdf/usage.html) and CDF (http://cdf.gsfc.nasa.gov).   In contrast, Quick Dataset is not intended to be a standard but is only used internally.   At present, mappings exist from NetCDF and CDF to QDataSet, and it is expected that in the future, users will have the option of exporting data into NetCDF; and CDF export is now possible.

Autoplot is distributed as a simple customizable software package that can be used in client or server modes.  Potential users of Autoplot include data providers who wish to quickly provide graphics for the data they serve instead of writing their own ad-hoc software.  Autoplot can be used in three different modes: (1) as a server-side application to produce static image files, (2) as a client-side ("thick client") application that is installed using Java's Webstart or by downloading a OS-specific installation package or (3) in the web browser ("thin client") as a Java applet.

Autoplot's architecture is such that the data access layer is separate from those for handling and viewing data, and it can be extended to allow access to new types of data. Similar to a web browser, the view that a user sees is internally represented in a way that is similar to that of the DOM (Document Object Model) recommendation of the W3C (World Wide Web Consortium; http://www.w3.org/DOM/) that is used by modern web browsers . Autoplot's DOM is a tree-like structure of nodes with properties that contain the entire application state, supporting undo and redo operations, and the data retrieved from external sources.  Its model-view-controller architecture simplifies the process of creating different "views."

This paper has three audiences.  Users who want to understand Autoplot's overall capabilities, data service providers who want to use it to provide access to their data, and software developers who want to understand its architecture.  Overall information about Autoplot's capabilities is provided in Section 2.  Section 3 contains information for data providers who would like to use Autoplot to provide a visualization service on their data.  Section 4 contains information about the overall software architecture.

# 2. Using Autoplot

The primary goal of Autoplot is to enable the plotting of data found on the web using a simple and familiar interface. Many of its design features are direct translations of features found in modern internet browsers. An address bar is used to enter the URI location of a dataset. URIs may be bookmarked, and bookmark lists can be exported and imported. Remote files are downloaded and cached locally in a directory named "autoplot_data" in the user's home directory. Currently, Autoplot can handle datasets served over the HTTP, FTP, and SFTP protocols. As shown in Figure 1, loaded data are displayed with axes settings, symbols, and plot style choices that were automatically selected by inspecting the data values as they were read (and by possibly using hints found in the metadata).

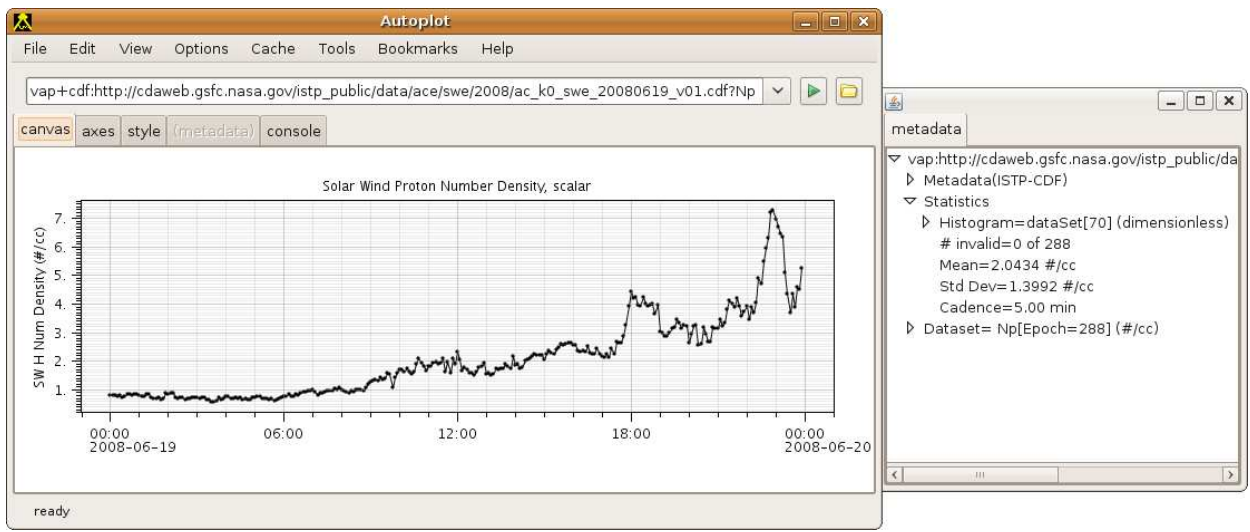

**Figure 1:** Autoplot's interface is similar to that of a web browser. The dataset location, or URI, is entered in the address bar and data are loaded and plotted with reasonable default settings. The metadata tab allows users to drill down into the parameter metadata, data set statistics, and the datum values. Axis and style settings tabs control labels and colors. The vap+cdf: prefix in the URI was inserted automatically by Autoplot to show how the data is being handled. This prefix notation identifies the Data Source Plug-in that has been used to interpret the URI. Data Source Plug-ins are described in Section 4.

Autoplot URIs provide a compact means for referring to data. They are intended to be stored in bookmarks, pasted into emails, and found in parameter lists. The URI refers to a specific dataset, which is a set of numbers and relationships between them, with associated metadata for their use. As described in Section 3.3, the same URI used to plot the data in Autoplot can be used with to pull the data directly into IDL and MATLAB.

**2.1 Accessing data**

Given a URI to a data source, Autoplot must decide on the appropriate file reader and metadata interpreter to use. Data Source Plug-ins provide access to each data type and are typically automatically chosen based on information in the URI, for example, by using the file name extension. These plug-ins are interfaces that translate a given data sources' data representation into that used by Autoplot. For example, the .dat extension resolves to the ASCII table Data Source Plug-in, which is given a URI and returns a QDataSet. Many Data Source Plug-ins are available, including ones for files containing ASCII tables along with standardized file or stream formats including CDF, Excel spreadsheets, NetCDF, OpenDAP, and CEF.

Often more than just a file name is required to identify a dataset at a high level of granularity. For example, a CDF file may contain many parameters. If only a single parameter is of interest, the parameter name can be specified in the URI as a query string, which is the same form that is used for passing parameters to a web application. When the following URI is entered, a plot of a single parameter is rendered:

```
http://autoplot.org/data/autoplot.cdf?BGSM.
```

When a URI that corresponds to many parameters (e.g., if the query string in the above URI was omitted), Autoplot will attempt to assist in correcting the URI by providing a drop-down list of parameters that it can plot (a "completion list" containing parameter names, in this case), or will start a GUI that allows the user to graphically specify what should be plotted. A completion list can be invoked manually with the keystroke combination Control-Space.

Autoplot can do client-side aggregation if the URI contains wild card expressions. For example entering,

```
http://cdaweb.gsfc.nasa.gov/istp_public/data/ace/
   swe/$Y/ac_k0_swe_$Y$m$d_v...cdf?Np&timerange=2008-June
```

will cause the files that match the template to be downloaded and combined together to form the complete time series of the parameter Np in June of 2008. The data required to create the complete time series is spread across files having names that include four-digit year ($Y), two-digit month ($m), and two-digit day ($d) that are stored in directories partitioned by year. The ".." following the "_v" in the filename matches any version number. An example plot of an aggregated time series is shown in Figure 2.

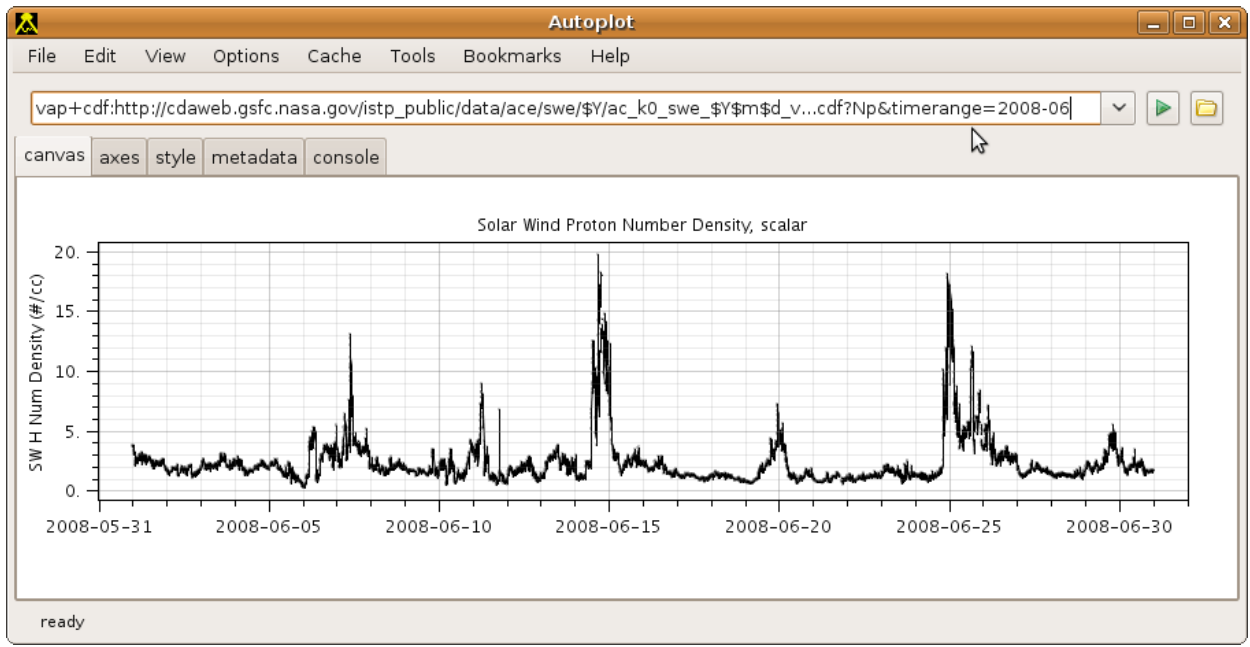

**Figure 2:** Autoplot's aggregation is used when the URI contains a template rather than the name of just one file. The user only specified the URI and then thirty data files containing data in June, 2008 were downloaded, combined, and plotted.

**2.2 Controls**

Once data is loaded and displayed with initial automatic settings, axis ranges and labels can be changed easily. The Das2 library (http://das2.org/) provides fluid mouse controls for browsing the plot. Standard interactions include zooming and panning. The entire plot may be zoomed and panned and the same interaction gesture may be applied individually to the X- or Y-axis. These interaction gestures also apply to the color bar axis on spectrograms. When the user right-clicks on a plot, a context menu provides options including "Reset Zoom" which will jump back to the original zoom level. A number of additional mouse actions, listed in the right-click context menu, allow for viewing specific data values, computing distances and slopes when the user draws a line, and listing the position where the mouse is clicked or located.

The "Axis" tab provides controls for setting axis ranges and labels. Ranges are parsed as a unit, so for example, the time range "June 2008" is a valid entry, as is the word "to" for delimiting minimum and maximum values. Plot titles and axis labels may be set, and Unicode characters can be used for math and Greek symbols. Labels may also contain format control codes that allow equations to be rendered.

The application state is continuously logged, and Edit→Undo (or control-Z) will return to the previous state, and Edit→Redo (or control-Y) will redo the last undo operation.

The state of Autoplot can be saved into a ".vap" file, which is an XML file containing the entire DOM tree and represents the comprehensive application state after any user interaction has taken place. The Autoplot DOM is a tree-like structure containing all the properties that configure the application. It has nodes for all the plots, plot elements, and data on the canvas. The canvas node contains nodes for the rows and columns that lay out the page. The DOM can be inspected directly (and manipulated) using "Edit→Edit DOM" on the menu bar.

By default, the data required to restore the view is not stored in the .vap file. If another user opens the .vap file, the data required to restore the application state will be downloaded automatically when the file is opened. There is an option to store the data in the XML file, which is reasonable when restoring the application state requires a small amount of data. In a future version, the data required to complete the view to be saved locally in a zip file, which will enable off-line usage.

## 2.3 Multi-Panel plots

The canvas can show any number of plots, typically in a horizontal stack dividing the canvas so that parameters can be compared on the same time axis. This is just one configuration, and in general plots can be laid out in multiple rows and columns.

Additional plots are added to the canvas by selecting "File→Add Plot". This brings up a dialog with a new address bar and buttons to overplot, plot below, or plot. ("Plot" is the default action that replaces the current plot.)

Each plot contains "plot elements". An element points to data and a plot, and has a render type (e.g. spectrogram, series for a line plot, scatter, etc.) that specifies how that data should be rendered. This allows for overplots with different datasets plotted on the same axes, and makes it easy to move individual elements around the page. For example, a vector quantity (such as a three-component magnetic field time series) is internally represented by three plot elements, one for each vector component. The user has control of the style of each plot element, and an individual element can be moved to a new plot. An invisible "parent" element controls all component elements at once, and is accessible via the "layout" tab. Plots and plot elements may be easily rearranged or deleted using the layout tab (enabled using Options→Enable Feature).

When there are multiple plots, the application "focus" state determines what the axis and style tabs control and what URI is shown in the address bar. The focus is set as plot elements are added and when a plot (or element within a plot) is clicked. When the focus changes, the selected plot or plot element will flash briefly.

As an example of controls and plot elements, consider a stack of plots that all share a common time axis (technically, their time axes elements are bound). A binding forces two properties of the DOM to be equal. Bindings can be added and removed at any time, and are themselves part of the DOM. This makes the configuration very flexible. For example, if we have two spacecraft in the solar wind, we can have a stack of panels from one spacecraft and then another stack from the other spacecraft with shifted time axes. Bindings are created automatically when a new plot appears to a time span similar to the application time range property. The user can explicitly create and delete bindings using the axes' context menus. Figure 3 shows a case where a stack of time series have bound time axes (top two) or shifted time axes (bottom). When the user modifies bottom axes in the stack, the other plots are modified to show the selected time range. If the user modifies the time range in one of the top two plots in the stack, the other plots are updated accordingly.

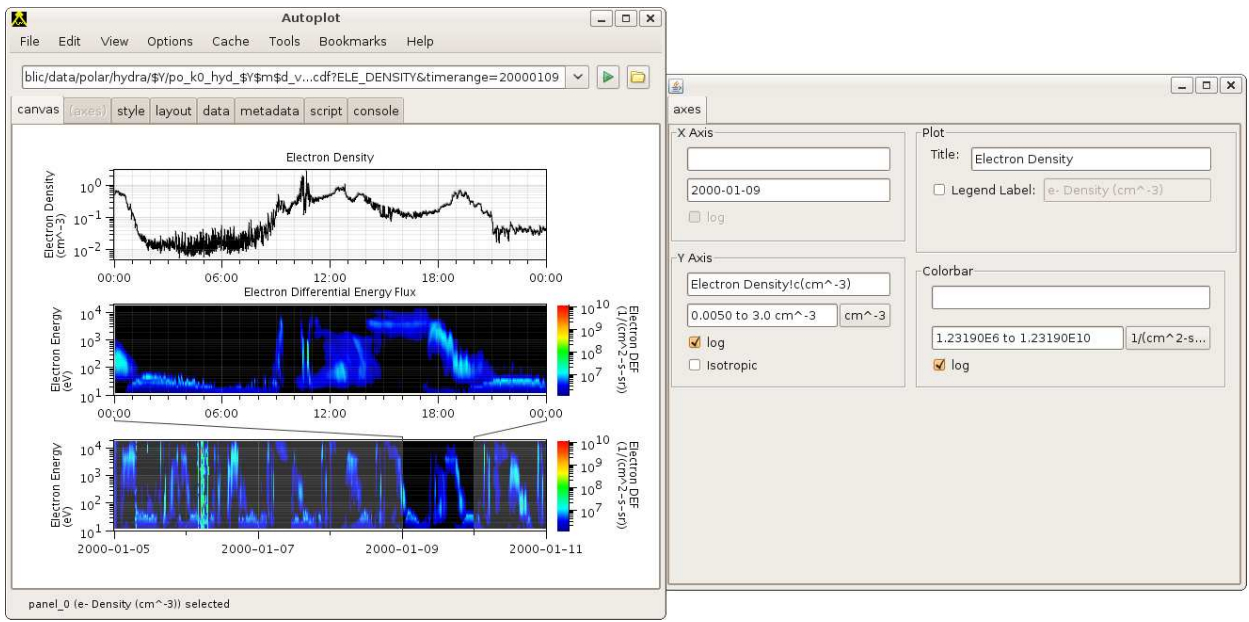

**Figure 3:** Example display in Autoplot with three plots. The top two plots have their X axes bound so they move together, and are a zoom-in of the bottom plot. This bottom plot is a "context overview" and dragging the window in the bottom panel controls the upper two panels. The axis tab, undocked from the application, can be used for setting the labels of the plot with focus.

## 2.4 Scripting

A rich library of operators that work on QDataSets has been developed. These include standard methods for reducing and filtering data. Metadata-aware arithmetic operators that manage fill values, physical units, and dependencies are also available.

Coupled with the Python scripting language and an API for Autoplot's data access libraries, a rich language for data processing is available. With a syntax and functions similar to IDL and Matlab, users can combine and process data to create ad-hoc datasets.

Given a script, the Jython Data Source Plug-in (Jython is a Java-based Python interpreter) returns a QDataSet. These scripts can be accessed remotely by way of the virtual file systems, allowing data processes to be published via HTTP.

Autoplot's view may also be manipulated with Jython program, in the same way that a web browser's view may be modified with a Javascript program. An example usage is the Jython program that creates the "PNG Walk Tool", which allows the user to flip through a set of images, shown in Figure 4.

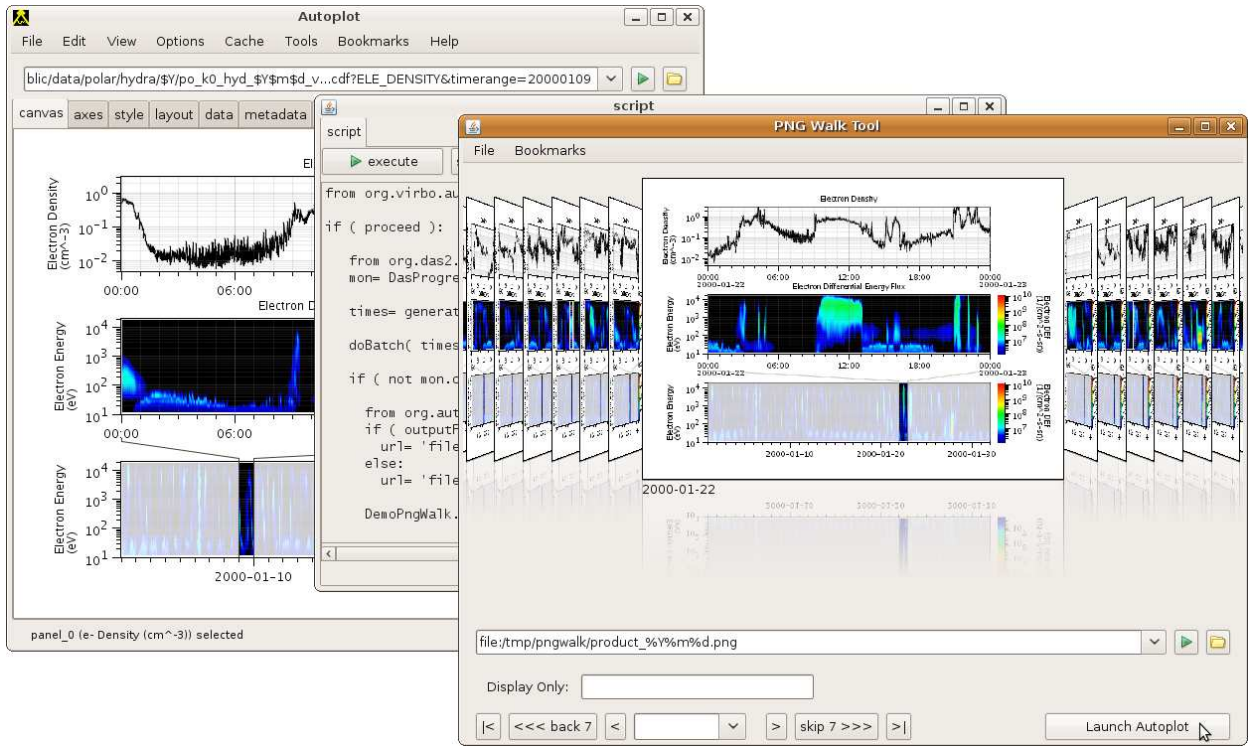

**Figure 4:** The "PNG Walk Tool" allows the user to quickly browse through a month's worth of data at a cadence of one PNG file per day. The PNG images were created by a script, written in Jython, and the "launch Autoplot" button will reset Autoplot to the image settings so the data can be explored interactively.

## 3. Use Cases

Autoplot was designed to have a clear separation between data accessors (Data Source Plug-ins, described in Section 4.1), its internal data model (QDataSet, described in Section 4.2), the application state (represented as a DOM), views of the application (the canvas), and interactions with or control of the DOM and canvas (the GUI). This allows us to easily use these pieces individually in other applications. In Section 2, we have described the full ("thick client") application, which uses all of these parts. In this section we give examples how different parts of Autoplot can be used.

### 3.1 Applet

Autoplot works as an unsigned (security restricted) Java applet. A single jar file is used with the HTML applet tag to embed Autoplot's canvas in a web page. The user does not see an address bar or the top menu bar (that contains the options of "File", "View", etc.). The size of the jar file is approximately 300Kb, which is much less than that of the full application, mostly because only a few of the Data Source Plug-ins work in this security-restricted mode and because the GUI is not provided. The applet uses Das2's canvas controls, so the user can interact and control the plot and plot elements. Instead of using a Java-based GUI to support modification of the canvas, a Javascript interface has been developed so that a user interface could be built using HTML-based GUI elements (for example, the address bar, tabs, and menu list could be implemented using Javascript and HTML).

### 3.2 Image service

Autoplot can be run on a Java-enabled web server to produce static graphics. In this case the user simply enters the URL to an Autoplot-enabled server with an Autoplot URI appended to it. The URI may contain optional parameters to set the labels and time ranges. The server simply returns a PNG, PDF, or SVG image.

### 3.3 Data reader for IDL and MATLAB

A user that does not want to use Autoplot's canvas or GUI may still use its Data Source Plug-ins to easily pull data specified by an Autoplot URI into an array in IDL or MATLAB. The following example puts the 287 values in a column named *data* located inside of an Excel spreadsheet file named *swe-np.xls* on a remote server into an IDL variable named *data*:

```
IDL> data= OBJ_NEW('IDLjavaObject$APDataSet', 'org.virbo.idlsupport.APDataSet')
IDL> data->setDataSetUrl, 'http://autoplot.org/data/swe-np.xls?column=data'
IDL> data->doGetDataSet
IDL> help, data->values()
<Expression> DOUBLE = Array[287]
```

The traditional way that an IDL or MATLAB user would extract these data is to download it to a local file and then use a vendor-supplied routine to read the data into an array. The disadvantage of this approach is that a different routine is required for each file type. If Autoplot is used, only a single routine is needed to read data from a variety of file types.

### 3.4 Translation service

Because Autoplot's data access libraries represent various forms of data using one common model, it can be used to create a reformatting and translation service.  Although many Data Source Plug-ins handle directory trees of files, Jython scripting can be used to handle data stored in more complex directory trees to extract, subset, and output specific parameters.  This is somewhat similar to the concept of "aggregation" used in NetCDF.

## 4. Architecture

Figure 5 shows the overall architecture of Autoplot.

*(Figure 5 is at end of document)*

**Figure 5:** Some of the components of Autoplot showing the path of a dataset URI and the QDataSet containing the data the URI represents.

### 4.1 Data Source Plug-ins

Autoplot can read data from many sources using "Data Source Plug-ins" that know how to access data and metadata from a data source.  Each data source plug-in has an associated list of extension types that they can handle.  Their basic function is to return a dataset given URI.  Data Source Plug-ins also provide URI validation and completion. (Completion defines the scheme for valid URIs, so for example, it identifies what parameters are found in the URI, and what the valid parameter values are; this is what determines the pop-up completion lists the user sees in the address bar.)

The data source can be explicitly specified by prefixing the URI with a scheme identifier of the form *vap+ext:,* where *ext* is the extension that identifies the data source. For example,

```
vap+dat:file:///home/user/myfile.asc?delim=comma&skip=5
```

indicates that the ASCII table data source plug-data (which is associated with the *dat* type) should be used to read the file. The query string following the question mark instruct this data source plug-in that that the column delimiter is a comma and it should skip the first 5 lines of the file.

The data source plug-in may also provide a GUI to creating a query string.  For example, if the user does not specify a query string, the ASCII data source plug-in will open an interactive GUI for identifying field delimiters and selecting columns to plot.  Similarly, the CDF data source plug-in shows the list of parameters found in the file along with metadata information associated with each parameter.

Another service some Data Source Plug-ins provide is the formating or exporting of data.  For file-based data sources this service would read a dataset and write it out into a particular form.

In general, the data source extension ID is automatically selected if one is not given, but it can be explicitly declared if, for example, the file name's extension is unconventional. Entering "about:plugins" in the address bar will show the list of registered extension ids. When the id is not explicitly stated, it is implicit from the extension of the filename found in the URI.

The metadata produced by the data source plug-in are returned as a tree-like structure of *name=value* pairs, and the data returned by a data source plug-in is a QDataSet.

**4.2 QDataSet**

QDataSet [Faden et al., 2009] (for "Quick Dataset") is the result of many iterations by the author to create a general model for representing many of the data structures encountered in space physics. It borrows conventions from CDF and NetCDF and has a simple Java interface; most of its specification in a semantic layer. The model is easily represented or mapped to data structures in other languages such as Python, MATLAB, IDL, and XML. The QDataSet data representation model was first developed for the IDL application PaPCo [Faden et al., 2010] and has matured as Autoplot incorporated more data types.

A QDataSet is a 0,1,2 or 3-index array with properties (name=value pairs) for metadata. There are familiar properties like UNITS ("microseconds since 2009-10-04T00:00"), TITLE, FILL_VALUE and VALID_MIN, but also new properties like BASIS ("since 2009-10-04T00:00") and CONTEXT_0 which retains contextual data that would otherwise be lost.

Some properties describe the use of an array index. For example, DEPEND_0 indicates the zeroth index is a physical dimension that depends on another dataset. The property value in this case is another QDataSet describing the independent parameter. BUNDLE_1 indicates that the index simply ties a number of datasets together that might have a common DEPEND_0. The property BINS_1 indicates the first index accesses boundaries within a physical dimension, like the minimum and maximum, specifying the interval over which a measurement was taken. The rank is the number of indices a dataset has, which is different from the number of physical dimensions it occupies (which is its dimensionality). Note that DEPEND increases the dimensionality by one, BUNDLE increases the dimensionality by the number of elements indexed, and BINS does not change dimensionality. Figure 6 shows three examples of QDataSets and their renderings.

*(Figure 6 is at end of document.)*

**Figure 6:** The left column shows three Autoplot canvases along with their associated QDataSet in the right column. Each box in the right column represents a QDataset, and has an identifier along with indices having an element count. Example properties of QDataset set are listed. The NAME, DEPEND_0, DEPEND_1, and BUNDLE_1 properties are indicated graphically next to their semantic meaning: NAME identifies a QDataset, and DEPEND and BUNDLE properties link QDatasets together to form more complex QDatasets.

QDatasets are nested, so one with one index (a "rank 1" QDataset) is an array of rank 0 QDatasets, and each index inherits context from the previous. Similarly, a rank 2 QDataset is an array of rank 1 QDatasets, and so on. Note that this model allows for attaching arbitrary metadata to any individual measurement, and allows for arrays of QDatasets with different lengths. We define slicing as extracting a QDataset from its context, which results in a rank reduction. QUBE is a special property to identify that the QDataset's indices can be exchanged, transposed, and that any index can be sliced.

## 5. Software Availability and Requirements

The Autoplot software is available at http://autoplot.org/, as are links to discussion groups, documentation, and source code. Its license is GPL 2. At the top of this page is a Java Web Start link that should download, install, and launch the application, provided that Java 5 is installed on the desktop. Applet and servlet versions are found on the web site, along with demonstrations of their use.

Autoplot uses a number of third-party libraries. For accessing NetCDF, OpenDAP, and HDF files, we use the netcdf-4.1 and dods-1.1.7 Java libraries. For CDF files, we use the binary library provided by NASA/SPDF via webstart. We also use Jython-2.2.1 to provide scripting, and the das2 library from the University of Iowa Radio and Plasma Wave Group.

## 6. Conclusions

Autoplot was started with the goal of providing a plotting tool for the Virtual Observatories that would be useful through out the community. By breaking it up into useful components, each providing useful services in software, this general purpose code can be extended and reused to create specialized applications. Autoplot will continue to be developed for a quickly growing set of applications, including closer integration with the Virtual Observatories and leveraging off the SPASE (Space Physics Search and Extract; http://spase-group.org/) metadata model to examine fully developed science data, but also low-level engineering displays. It is at the center of a growing community of scientists with various science applications.

Autoplot has proven to be a flexible tool for exploring data and accessing existing data resources found on the web. Though developed for use in the Virtual Observatories with magnetospheric data, its flexible data representation model should make it useful for data in many of the Earth sciences.

# 7. Acknowledgements

Autoplot development was supported by NASA grant NNX07AB70G (VxO for S3C Data: The Virtual Radiation Belt Observatory), and NASA Grant No. NNX07AC95G (Virtual Magnetospheric Observatory) issued through the Virtual Observatories for Solar and Space Physics Data (S3CVO). Additional support for development has been provided by the Los Alamos National Laboratory Radiation Belt Storm Probes project. We also acknowledge Edward Jackson, Edward West, and Larry Granroth for creative contributions and consulting, and the Plasma Wave Group at the University of Iowa for development support and the use of its open-source Das2 library, and the many early adopters who provided feedback.

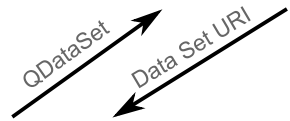

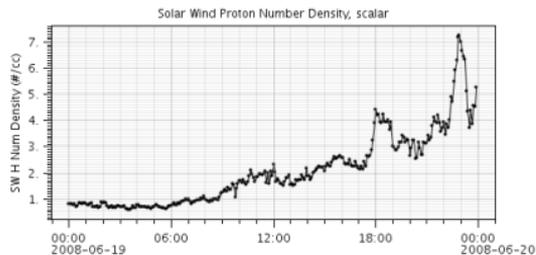

```
Np[Epoch=288]
LABEL="SW H Num Density"
UNITS=#/cc
```
↓ DEPEND_0
```
Epoch[288]
UNITS=milliseconds since 1970-01-01T00:00
```

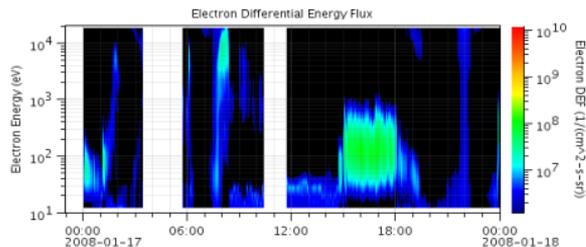

```
Energy[29]
UNITS=eV
SCALE_TYPE=log
LABEL="Electron Energy"
```
← DEPEND_1
```
Electron_DEF[Epoch=6260,Energy=29]
LABEL="Electron DEF"
TITLE="Election Differential Energy Flux"
UNITS=1/(cm^2-s-sr)
VALID_MIN=0.0
VALID_MAX=1.24e10
QUBE=true
```
↓ DEPEND_0
```
Epoch[6260]
UNITS=milliseconds since 1970-01-01T00:00
```

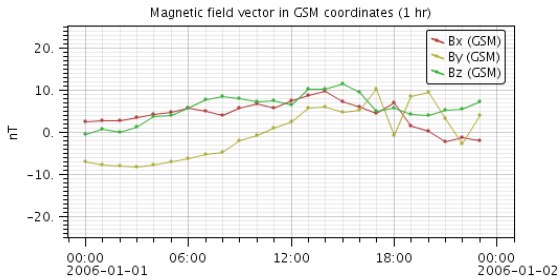

```
Bx
LABEL="Bx (GSM)"
By
LABEL="By (GSM)"
Bz
LABEL="Bz (GSM)"
```
← BUNDLE_1
```
BGSM[Time=24,[Bx,By,Bz]]
UNITS=nT
VALID_MIN=-65534.0
VALID_MAX=65534.0
QUBE=true
```
↓ DEPEND_0
```
Time[24]
UNITS=days since 2000-01-01T00:00
```